\documentclass[prd,preprint,aps,amsmath,amssymb,showpacs,nofootinbib,11pt]{revtex4-2}

\usepackage{graphicx} 
\usepackage{float}
\usepackage{color}
\usepackage[normalem]{ulem}
\usepackage[colorlinks=true,citecolor=blue,linkcolor=blue,urlcolor=blue]{hyperref}

\begin{document}
\count\footins = 1000

\title{Addressing Local Realism through Bell Tests at Colliders}

\author{Matthew Low}
\email{matthew.w.low@pitt.edu}
\affiliation{Pittsburgh Particle Physics, Astrophysics, and Cosmology Center,\\Department of Physics and Astronomy, University of Pittsburgh, Pittsburgh, USA}

\date{\today}

\begin{abstract}
One of the most notable aspects of quantum systems is that their components can exhibit correlations much stronger than those allowed by classical physics.  Two examples of quantum correlations are quantum entanglement and Bell nonlocality, but generally there is a hierarchy of many types of quantum correlations.  Among these correlations, Bell nonlocality holds a special place because it plays a dual role in distinguishing theories where local realism is a valid description.  A Bell test, which is a test of local realism, typically needs to be augmented with assumptions to address possible loopholes in the experimental setup.  In this work, we study Bell tests in experiments in which the detector reports the correct outcome with a specified probability.  This mirrors the situation at high-energy colliders, where particle spins are not measured directly but inferred from the angular distributions of their decay products.  We show that, in this setup, a test of local realism is not possible.  Quantum correlations, however, are still present, measurable, and informative in high-energy colliders.  These correlations are the building blocks of the interesting, developing quantum information science program at high-energy colliders.  The measurements of entanglement by the ATLAS and CMS experiments are the first steps in this initiative.
\end{abstract}

\maketitle
\newpage
\tableofcontents

\section{Introduction}
\label{sec:introduction}

Quantum mechanics underpins much of modern physics; however, its nature is still mysterious.  The evidence in favor of quantum mechanics is enormous and supported by generations of experiments~\cite{Freedman:1972zza,Aspect:1981zz,Aspect:1981nv,Aspect:1982fx,Bouwmeester:1997slj,Giustina:2015yza}.  Traditional quantum mechanical experiments are performed at low energies in systems of photons or electrons.  These systems may consist of two or more particles that exhibit correlations that exceed what is possible from classical physics.  A single particle represents a single unit of information, the bit, or for elementary particles whose interactions are governed by quantum mechanics, the qubit~\cite{Schumacher:1995nrx}.

Different quantum mechanical quantities characterize different properties of systems.  For example, quantum entanglement classifies states as entangled, meaning that one system cannot be fully described without also describing a second system, or as separable, meaning that the dynamics of the systems factorize~\cite{Einstein:1935rr,Horodecki:2009zz}.  Generally, there is a hierarchy of these classifications that separates possible states into sets.  Some known correlations that straightforwardly fit in this hierarchy are quantum discord~\cite{Zurek_2000,Ollivier:2001fdq}, quantum entanglement~\cite{Einstein:1935rr,Horodecki:2009zz}, steerability~\cite{PhysRevLett.98.140402}, Bell nonlocality~\cite{Bell:1964kc}, and non-negative conditional entropy~\cite{Horodecki_2005}.  Colloquially, correlations corresponding to smaller sets are referred to as stronger correlations.

Among these correlations, Bell nonlocality holds a special place because it plays a dual role as a quantum correlation and as a test that excludes local realism as a description of nature.  We use the term Bell nonlocality to refer to the quantum correlation that classifies a quantum state and reserve the term Bell test to refer to the test of local realism.  In this work, we discuss whether Bell tests can be performed at colliders.

Recently, collider experiments have begun probing quantum phenomena directly~\cite{Afik:2025ejh}.  This started with quantum entanglement between spins in the $t\bar{t}$ system~\cite{Afik:2020onf}, which has already been measured by ATLAS~\cite{ATLAS:2023fsd} and CMS~\cite{CMS:2024pts,CMS:2024zkc}.  Following this, early proposals for measuring Bell nonlocality, which may be measurable in $t\bar{t}$ at the high luminosity run of the Large Hadron Collider (LHC), have been put forward~\cite{Fabbrichesi:2021npl,Severi:2021cnj,Afik:2022kwm,Aguilar-Saavedra:2022uye}.  Other quantum correlations like quantum discord~\cite{Han:2024ugl}, steerability~\cite{Afik:2022dgh}, non-negative conditional entropy~\cite{Han:2024ugl}, and the nonlocal advantage of quantum coherence~\cite{Rai:2025qke} have also been studied in the $t\bar{t}$ final state.  A variety of other final states and other high-energy colliders have also been suggested as systems in which to study Bell nonlocality~\cite{Barr:2021zcp,Fabbrichesi:2022ovb,Aguilar-Saavedra:2022wam,Ashby-Pickering:2022umy,Altakach:2022ywa,Fabbrichesi:2023cev,Dong:2023xiw,Morales:2023gow,Aoude:2023hxv,Bernal:2023ruk,Bi:2023uop,Ma:2023yvd,Han:2023fci,Cheng:2023qmz,Ehataht:2023zzt,Maltoni:2024tul,Barr:2024djo,Morales:2024jhj,Maltoni:2024csn,Afik:2024uif,Wu:2024asu,Cheng:2024btk,Demina:2024dst,LoChiatto:2024dmx,Gabrielli:2024kbz,Cheng:2024rxi,Wu:2024ovc,Gao:2024leu,Cheng:2025cuv,Fabbrichesi:2025ywl,Han:2025ewp,Varma:2025gkp,Horodecki:2025tpn,Fabbrichesi:2025ifv,Zhang:2025mmm,DelGratta:2025qyp,Ding:2025mzj,Fabbrichesi:2025rsg,Rai:2025qke,Afik:2025grr,Qi:2025onf,Cheng:2025zcf,Grossi:2024jae}.  In this work, we exclusively discuss spins as qubits, however, this is an analogous situation when using flavor quantum numbers as qubits.  In this case, the different decay times neutral meson systems allow different correlations to be probed.  Unfortunately, local realism cannot be tested using flavor in current experiments~\cite{Bertlmann:2004cr,Bramon:2004pp}.

Although early work has investigated the applicability of Bell tests at colliders~\cite{Abel:1992kz,Dreiner:1992gt,Li:2024luk,Fabbrichesi:2025aqp,Bechtle:2025ugc,Abel:2025skj}, some confusion persists.  In this work we confirm the statement of Refs.~\cite{Abel:1992kz,Dreiner:1992gt,Bechtle:2025ugc,Abel:2025skj} that local realism cannot be tested at colliders.  We show that local realism cannot be tested with current high-energy particle detectors for two primary reasons.  Firstly, there are no independent detector settings, which are required for Bell tests.  As a result, Bell's inequality reduces to differential cross sections, characterized by commuting measurements, which can be described by a local hidden-variable model.  Secondly, even if we had independent detector settings, we do not measure the spins of particles directly.  The rest-frame decay-product angular distributions can be used to infer the polarizations and spin correlations of particles, but the inherent distribution of these angles is too wide to be used as a proxy for the spins.  Consequently, while strong spin correlations may exist in the underlying quantum state, they are diluted when comparing the correlations between rest-frame decay angles.  As a result, Bell's inequality cannot be violated in this system.  We demonstrate this with a direct calculation.

The organization of the paper is as follows.  In Sec.~\ref{sec:Bell} we present several variations of Bell's inequality for two qubits and discuss the role of loopholes and assumptions.  Section~\ref{sec:detection} contrasts how a Bell test is performed with an ideal detector and with an imperfect detector.  In Sec.~\ref{sec:colliders} we apply these situations to high-energy colliders where assuming underlying spins is analogous to measuring with an ideal detector and where using the rest-frame decay-product angles without assumptions is analogous to using an imperfect detector. We briefly address the differences between Bell nonlocality and entanglement in Sec.~\ref{sec:entanglement}. Sec.~\ref{sec:conclusions} contains our conclusions and outlook.

\section{Bell Tests}
\label{sec:Bell}

The Einstein-Podolsky-Rosen (EPR) paradox~\cite{Einstein:1935rr} asserts that quantum mechanics is not a complete description of nature if local realism is assumed.  Under local realism, both the position and the momentum of a particle can be known, both of which are not simultaneously predicted from quantum mechanics. Bohm reformulated this paradox using the spin along two different axes as complementary variables rather than momentum and position~\cite{Bohm:1951xw,Bohm:1951xx,Bohm:1957zz}.

In 1964 Bell suggested a gedanken experiment to address the EPR paradox directly~\cite{Bell:1964kc}.  Bell proposed a setup in which two photons are emitted from a common source.  One photon is detected by Alice, who can detect the polarization of the photon given a detector setting $\alpha$, which in this case is given by the angle of the polarizer.  Bob detects the second photon, at a spatially-separated location, with a separate detector setting choice $\beta$.  The outcome of a single instance of the experiment is characterized by Alice's measurement outcome $o_A$ and Bob's measurement outcome $o_B$, given their respective detector settings, $\alpha$ and $\beta$.  The results are governed by a probability distribution $P(o_A o_B | \alpha \beta)$ which means that outcomes can vary in each instance of the experiment.
  
The experiment must, therefore, be performed many times to obtain an estimate of the probabilities.  In Bell's experiment the outcome of Alice's measurement is $+1$ or $-1$, but we will often use the shorthand $o_A \in \{ +, - \}$, and likewise for Bob $o_B \in \{ +, - \}$, leading to the joint outcomes of $\{ ++, +-, -+, -- \}$.  Let the number of instances measured with the outcome $o_A$ and  $o_B$ with detector settings $\alpha$ and $\beta$ be $N_{o_A o_B}(\alpha,\beta)$.  A single run of the experiment then increases the count of one of the following: $N_{++}(\alpha,\beta)$, $N_{+-}(\alpha,\beta)$, $N_{-+}(\alpha,\beta)$, or $N_{--}(\alpha,\beta)$.

The expectation value of the outcome of the experiment in the configuration of $(\alpha,\beta)$ is given by $E(\alpha,\beta)$ where
\begin{equation}
\label{eq:exp-value-exp}
E(\alpha,\beta) = 
\frac{N_{++}(\alpha,\beta) + N_{--}(\alpha,\beta) 
- N_{+-}(\alpha,\beta) - N_{-+}(\alpha,\beta)}
{N_{++}(\alpha,\beta) + N_{--}(\alpha,\beta) 
+ N_{+-}(\alpha,\beta) + N_{-+}(\alpha,\beta)},
\end{equation}
which ranges from $-1 \leq E(\alpha,\beta) \leq 1$.  Eq.~\eqref{eq:exp-value-exp} can be measured in an experiment and makes no reference to an underlying theory description.

On the other hand, to predict the value of $E(\alpha,\beta)$ given a theory, one uses the predicted probability distribution $P(o_A o_B | \alpha \beta)$ and calculates
\begin{equation}
\label{eq:exp-value-th}
E(\alpha,\beta) = \sum_{o_A, o_B} (o_A o_B) P(o_A o_B | \alpha \beta).
\end{equation}
This prediction may differ for different underlying theories.

Bell proposed that Alice switches between two detector settings $\alpha$ and $\gamma$ and Bob switches between two detector settings $\beta$ and $\gamma$.  Assuming that the two photons are produced in a singlet state, equivalent to the assumption that $E(\gamma,\gamma)=-1$, Bell derived the inequality~\cite{Bell:1964kc}
\begin{equation}
\label{eq:bell-original}
|E(\gamma,\alpha) - E(\gamma,\beta)| \leq 1 + E(\alpha,\beta).
\end{equation}
The verification of this inequality consists of performing the experiment a sufficiently large number of times to obtain estimates of the expectation values in three different configurations of detector settings: $\gamma$ and $\alpha$, $\gamma$ and $\beta$, and $\alpha$ and $\beta$.  All theories obeying local realism, including theories with hidden variables, obey Eq.~\eqref{eq:bell-original}.  An experiment that shows the violation of Bell's inequality would resolve the EPR paradox by indicating that the assumption of applying local realism to quantum mechanics is incorrect.

Locality, in this context, heuristically, means that the outcome of Alice's measurement does not depend on Bob's measurement and vice versa.  More formally, it can be stated as follows, adapting the presentation in Ref.~\cite{Brunner:2013est}.

The probability of Alice's measurement is $P(o_A | \alpha)$ and the probability of Bob's measurement is $P(o_B | \beta)$.  The probability of the joint outcome is $P(o_A o_B | \alpha \beta)$ and generally due to correlations, we have $P(o_A o_B | \alpha \beta) \neq P(o_A | \alpha) P(o_B | \beta)$.  Locality implies that any correlations between Alice’s and Bob’s outcomes arise from shared hidden variables $\lambda$ determined at the source and carried to the detectors.  In this way $P(o_A o_B | \alpha \beta, \lambda) = P(o_A | \alpha, \lambda) P(o_B | \beta, \lambda)$.  Integrating over the possible values of $\lambda$ leads to
\begin{equation}
\label{eq:locality}
P(o_A o_B | \alpha \beta) = \int d\lambda \: q(\lambda) P(o_A | \alpha, \lambda) P(o_B | \beta, \lambda),
\end{equation}
where $q(\lambda)$ is the distribution of $\lambda$.  The form $q(\lambda)$, independent of $\alpha$ and $\beta$, encodes the assumption that detector settings are independent of $\lambda$.  Eq.~\eqref{eq:locality} is a statement of the locality condition.

Realism assumes that each photon possesses definite values for all observables, independent of measurement.  For instance, realism asserts that in a single instance of the experiment the polarization of a photon along any axis, such as $\hat{x}$, $\hat{y}$, and $\hat{z}$, has a definite value.

In 1969, Clauser, Horne, Shimony, and Holt dropped the requirement that the particles originate in a singlet state at the cost of increasing the detector configurations from three to four.  Their inequality, called the CHSH inequality~\cite{Clauser:1969ny}, is
\begin{equation}
\label{eq:CHSH}
|E(\alpha_1,\beta_1) 
- E(\alpha_1,\beta_2) 
+ E(\alpha_2,\beta_1) 
+ E(\alpha_2,\beta_2)| \leq 2.
\end{equation}
A key goal of Ref.~\cite{Clauser:1969ny} was to devise a practical version of Bell's gedanken experiment.  Two limitations present at the time were $(i)$ current detectors could only measure one value of the outcome $o_{A,B}$ in a single setup rather than both values and $(ii)$ the efficiency of detectors was low.  Together, these constraints made using Eq.~\eqref{eq:CHSH} difficult to implement.

When only a single value of the outcome $o_{A,B}$ is observable in a single run, the logical encoding of the experiment is that $o_{A,B} = +$ corresponds to the observation of the specified outcome (which could be $+$ or could be $-$) and $o_{A,B} = -$ corresponds to the non-observation.  While valid, this encoding renders Eq.~\eqref{eq:CHSH} infeasible to test because the detection efficiency multiplies all terms on the left-hand side making it impossible to observe a violation of Bell's inequality.

In the same work~\cite{Clauser:1969ny}, CHSH resolved this issue by building an inequality from the event rates, rather than expectation values
\begin{subequations}
\label{eq:rate-values}
\begin{align}
R(\alpha,\beta) &= \frac{N_{++}(\alpha,\beta)}{N_{\infty\infty}(\alpha,\beta)},\\
R(\alpha,\infty) &= \frac{N_{+\infty}(\alpha,\beta)}{N_{\infty\infty}(\alpha,\beta)},\\
R(\infty,\beta) &= \frac{N_{\infty +}(\alpha,\beta)}{N_{\infty\infty}(\alpha,\beta)}.
\end{align}
\end{subequations}
The notation $\infty$ indicates that both outcomes are included or in the low-energy experimental context it corresponds to removing the polarizer from that detector.  Eq.~\eqref{eq:CHSH} can then be expressed as~\cite{Clauser:1974tg}
\begin{equation}
\label{eq:CH1974}
-1 \leq R(\alpha_1, \beta_1) - R(\alpha_1,\beta_2) + R(\alpha_2,\beta_1) + R(\alpha_2,\beta_2) - R(\alpha_2,\infty) - R(\infty,\beta_1) \leq 0,
\end{equation}
which is called the single-channel CHSH inequality or the CH inequality.  In contrast, the CHSH inequality of Eq.~\eqref{eq:CHSH} is a dual-channel inequality.  The single-channel version of the gedanken experiment required seven different detector configurations, but was testable in experiments of the time.

The theory prediction of the rates is calculated from
\begin{subequations}
\label{eq:rate-values-th}
\begin{align}
R(\alpha,\beta) &= P(++ | \alpha \beta), \\
R(\alpha,\infty) &=  P(+ + | \alpha \beta) + P(+ - | \alpha \beta), \\
R(\infty,\beta) &=  P(+ + | \alpha \beta) + P(- + | \alpha \beta).
\end{align}
\end{subequations}
The first experimental observation of Bell inequality violation was carried out by Clauser and Freedman in 1972 using Eq.~\eqref{eq:CH1974}~\cite{Freedman:1972zza}.  Aspect and collaborators conducted a series of three experiments from 1981 to 1982~\cite{Aspect:1981zz,Aspect:1981nv,Aspect:1982fx}.  In Ref.~\cite{Aspect:1981zz} Aspect measured Eq.~\eqref{eq:CH1974}, in Ref.~\cite{Aspect:1981nv} Aspect measured Eq.~\eqref{eq:CHSH} for the first time, and in Ref.~\cite{Aspect:1982fx} Aspect measured Eq.~\eqref{eq:CH1974} but where the detector settings for one detector are chosen quasi-randomly between $\alpha_1$ and $\alpha_2$ and quasi-randomly between $\beta_1$ and $\beta_2$ for the other detector.  This was the first experiment to address the locality loophole.

A loophole is a scheme by which local realism may survive as a description of nature after the results of a given experiment~\cite{Larsson:2014rzp}.  Equivalently, a loophole specifies a class of local hidden variable models that are not excluded by the experiment in question.  An experiment with fewer loopholes excludes a larger class of local hidden variable models while an experiment with many loopholes excludes a smaller class of local hidden variable models.  Loopholes can either be closed by altering the experimental setup to address it or be ignored by invoking an assumption that local hidden variable models exploiting said loophole do not exist.

For example, the detection loophole states that if too many events go undetected, the remaining sample may be biased, potentially leading to a false violation of Bell's inequality.  An experiment with a low detection efficiency, therefore, does not exclude the class of local hidden variable models that cause a biased sample of events to be detected.  This loophole is addressed by designing an experiment with a higher detection efficiency or by invoking the fair sampling assumption~\cite{Garg:1986wd}.  This assumption states that the detected events are statistically representative of the total set of events.  

It is not possible to close every loophole, however, as the number of remaining loopholes shrinks, the class of allowed local hidden variable models becomes increasingly baroque~\cite{Larsson:2014rzp}.  Superdeterminism is a loophole that always persists.  For that reason, while it is preferable that an experiment address more loopholes, experiments which cannot address some loopholes, like collider experiments, are still interesting.

\section{Detection Scenarios}
\label{sec:detection}

In discussing detection, it is helpful to distinguish between the underlying outcome, the eigenvalue of the quantum state, and the reported outcome, the value actually recorded by the experimenter.

For an ideal detector, as in Sec.~\ref{sec:ideal}, there is no distinction between the underlying outcome and the reported outcome.  Explicitly, a measured outcome of $+1$ leads to a reported outcome of $+1$ and a measured outcome of $-1$ leads to a reported outcome of $-1$.

For an imperfect detector, as in Sec.~\ref{sec:imperfect}, the underlying outcome and the reported outcome can differ.  We consider the case where the underlying outcome is not measured directly, but rather is a continuous variable that is correlated with the underlying outcome is measured.  The continuous variable is then converted into a reported outcome via a chosen encoding.  As applied to collider experiments, like in Sec.~\ref{sec:colliders}, the continuous variable is the rest-frame angle of a decay product.  Given an encoding, an underlying outcome of $+1$ will lead to a reported outcome of $+1$ with a probability of $P_+$ and an underlying outcome of $-1$ will lead to a reported outcome of $-1$ with a probability of $P_-$.

These probabilities describe the relationship between the underlying and reported outcomes and are calculated by considering a measurement operator $\mathcal{O}_{\hat{n}}$ that measures a qubit along the $\hat{n}$-axis acting on an eigenstate $\rho_{\pm \hat{n}}$ of the measurement direction $\pm \hat{n}$.  For a single qubit the eigenstate is
\begin{equation}
\label{eq:rho-eigen}
\rho_{\pm \hat{n}} = \frac{1}{2}\left( 
\mathbb{I}_2 
\pm \hat{n} \cdot \vec{\sigma} 
 \right).
\end{equation}
Given a measurement operator, the relationship between the underlying outcome and the reported outcome is
\begin{equation}
\label{eq:exp-val-eigen}
\langle \mathcal{O}_{\hat{n}} \rangle_{\pm \hat{n}}
= {\rm tr}\Bigl[
\mathcal{O}_{\hat{n}} \rho_{\pm \hat{n}}
\Bigr].
\end{equation}
This relation between the outcomes is trivial for the ideal detector but becomes non-trivial for the imperfect detector and in collider experiments.

\subsection{An Ideal Detector}
\label{sec:ideal}

\begin{figure} 
  \centering
  \includegraphics[width=0.4\textwidth]{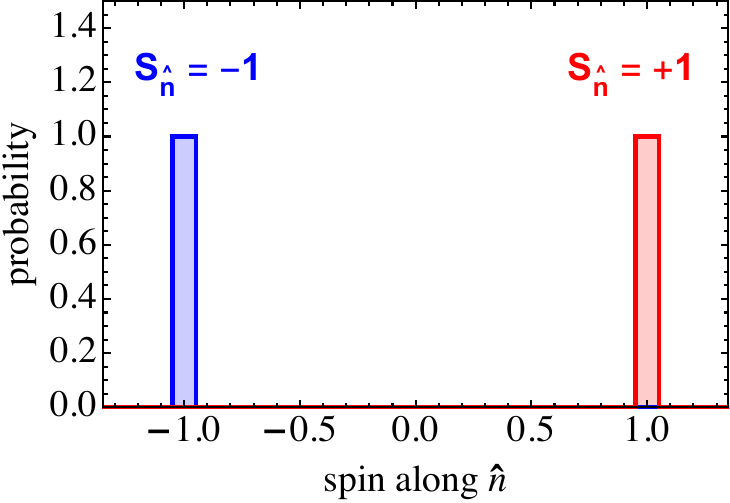}
  \caption{Distribution of reported outcomes from an ideal detector for spin, measured along the axis $\hat{n}$, for underlying outcomes $S_{\hat{n}}$ of $+1$ (red) and $-1$ (blue).}
\label{fig:det-ideal}
\end{figure}

An ideal detector is defined such that the underlying outcome and the reported outcomes are the same.  Informally, we would say that the spins are measured directly.  For such a measurement the operator is
\begin{equation}
\label{eq:opM-ideal}
\mathcal{O}_{\hat{n}} = \hat{n} \cdot \vec{\sigma},
\end{equation}
which leads to 
\begin{equation}
\label{eq:exp-val-eigen-ideal}
\langle \mathcal{O}_{\hat{n}} \rangle_{\pm \hat{n}} = {\rm tr}\Bigl[
(\hat{n} \cdot \vec{\sigma})
\rho_{\pm \hat{n}}
\Bigr]
= \pm 1.
\end{equation}
This is seen in Fig.~\ref{fig:det-ideal} which shows the relationship between reported outcomes and underlying outcomes, in the form of a distribution.\footnote{If the detector is perfectly ideal, as shown in Fig.~\ref{fig:det-ideal}, then a single instance of the experiment is sufficient to characterize the distribution of detector outcomes.  Such a situation is not required for an observation of Bell nonlocality.}

The general parametrization of a one-qubit system is
\begin{equation}
\label{eq:bloch-vector}
\rho = \frac{1}{2}
\Bigl(
\mathbb{I}_2 + \sum_i B_i \sigma_i
\Bigr),
\end{equation}
where the $B_i$, or in vector form $\hat{B}$, describe the polarization of the qubit and are collectively called the Bloch vector.

The expectation value of measuring this qubit along the $\hat{n}$-axis is
\begin{equation}
\label{eq:exp-val-bloch}
\langle \mathcal{O}_{\hat{n}} \rangle = {\rm tr}\Bigl[
(\hat{n} \cdot \vec{\sigma})
\rho
\Bigr]
= \hat{n} \cdot \hat{B}.
\end{equation}
The Fano-Bloch decomposition~\cite{Fano:1983zz} provides the general parametrization of a two-qubit state and is
\begin{equation}
\label{eq:fano-bloch}
\rho = \frac{1}{4}\Bigl( 
\mathbb{I}_2 \otimes \mathbb{I}_2 
+ \sum_i B^+_i \sigma_i \otimes \mathbb{I}_2 
+ \sum_j B^-_j \mathbb{I}_2 \otimes \sigma_j
+ \sum_{ij} C_{ij} \sigma_i \otimes \sigma_j
\Bigr),
\end{equation}
where $B^+_i$ parameterizes the polarization of Alice's qubit, $B^-_j$ parameterizes the polarization of Bob's qubit, and $C_{ij}$ parameterizes the spin correlations, and are collectively called the Fano coefficients.

The expectation value of measuring the first qubit along the axis $\hat{\alpha}$ and the second qubit along the axis $\hat{\beta}$ is
\begin{equation}
\label{eq:exp-val-fano}
\langle \mathcal{O}_{\hat{\alpha}} \otimes \mathcal{O}_{\hat{\beta}} \rangle = 
{\rm tr} \Bigl[ \bigl( \mathcal{O}_{\hat{\alpha}} \otimes \mathcal{O}_{\hat{\beta}}  \bigr)
\: \rho \:
\Bigr]
=
\hat{\alpha} \cdot C \cdot \hat{\beta}.
\end{equation}
The terms in the CHSH inequality, from Eq.~\eqref{eq:CHSH}, can be measured directly by the appropriate choices of $\hat{\alpha}$ and $\hat{\beta}$ resulting in
\begin{equation}
\label{eq:chsh-spins}
\left|
\vec{\alpha}_1 \cdot C \cdot \vec{\beta}_1
- \vec{\alpha}_1 \cdot C \cdot \vec{\beta}_2
+ \vec{\alpha}_2 \cdot C \cdot \vec{\beta}_1
+ \vec{\alpha}_1 \cdot C \cdot \vec{\beta}_2 
\right|
\leq 2.
\end{equation}
This depends on the four choices $\vec{\alpha}_1$, $\vec{\alpha}_2$, $\vec{\beta}_1$, and $\vec{\beta}_2$.  For a quantum mechanical system, the optimal choice of axes is known to lead to a value of the left-hand side of $2\sqrt{m_1 + m_2}$ where $m_1$ and $m_2$ are the two largest eigenvalues of the matrix $M = C^T \cdot C$~\cite{Horodecki:1995nsk}.  For a Bell state, this leads to a value of $2\sqrt{2}$, which coincides with the maximal value possible in quantum mechanics~\cite{Cirelson:1980ry}.

\subsection{An Imperfect Detector}
\label{sec:imperfect}

\subsubsection{A Simple Example}
\label{sec:imperfect-simple}

\begin{figure} 
  \centering
  \includegraphics[width=0.4\textwidth]{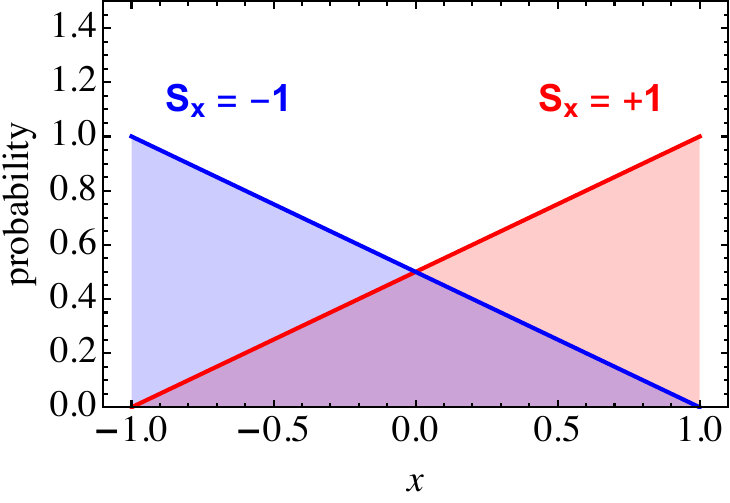}
  \caption{Distribution of reported outcomes from an imperfect detector for spin, measured along the $x$-axis, for underlying outcomes $S_x$ of  $+1$ (red) and $-1$ (blue).}
\label{fig:detection-imperfect-simple}
\end{figure}

In the case of an imperfect detector rather than measuring an eigenvalue of the spin operator, leading to a binary measurement of $+1$ or $-1$, a continuous value of $\vec{x}$ is measured.  In this section, we consider the simple one-dimensional case and the measurement operator
\begin{equation}
\label{eq:opM-simple}
\mathcal{O}(x) = \frac{1}{2} \left(\mathbb{I}_2 + x \sigma_x \right).
\end{equation}
The result of this measurement on an eigenstate $\rho_{\pm x}$ is a function of $x$
\begin{equation}
\label{eq:exp-val-eigen-simple}
\langle \mathcal{O}(x) \rangle_{\pm x}
= {\rm tr}\Bigl[ \frac{1}{2} \left(\mathbb{I}_2 + x \sigma_x \right)  \rho_{\pm x} \Bigr]
= \frac{1}{2}\left( 1 \pm x \right).
\end{equation}
This relationship between the underlying outcome and the reported outcome is now a distribution and is shown in Fig.~\ref{fig:detection-imperfect-simple}. 

The expectation value, in terms of the Bloch vector of Eq.~\eqref{eq:bloch-vector}, is
\begin{equation}
\label{eq:exp-val-simple}
\langle \mathcal{O}(x) \rangle = {\rm tr}\Bigl[
\frac{1}{2} ( \mathbb{I}_2 + x \sigma_x)
\rho
\Bigr]
= \frac{1}{2}\left( 1 + x B_x \right).
\end{equation}
and the expectation value, in terms of the Fano coefficients of Eq.~\eqref{eq:fano-bloch}, is
\begin{equation}
\label{eq:exp-val-simple-twoqubits}
\langle \mathcal{O}(x_A) \otimes \mathcal{O}(x_B) \rangle
=
\frac{1}{4}\Bigl(
1 + x_A B^+_x + x_B B^-_x + x_A C_{xx} x_B \Bigr).
\end{equation}
Unlike the previous case, this cannot be directly applied to the CHSH inequality.  Instead, an encoding must be specified that defines what constitutes a detected $+1$ and what constitutes a detected $-1$.  To extract binary outcomes from this continuous measurement, we define an encoding that assigns $+1$ to events with $x > 0$ and $-1$ to events with $x<0$. The resulting counts are:
\begin{subequations}
\label{eq:encoding-simple-gen}
\begin{align}
N_{++} &= 
N(x_A > 0, x_B > 0)
= \int_0^1 dx_A \int_0^1 dx_B 
\langle \mathcal{O}(x_A) \otimes \mathcal{O}(x_B) \rangle , \\
N_{+-} &= 
 N(x_A > 0, x_B < 0)
= \int_0^1 dx_A \int_{-1}^0 dx_B 
\langle \mathcal{O}(x_A) \otimes \mathcal{O}(x_B) \rangle, \\
N_{-+} &= 
N(x_A < 0, x_B > 0)
= \int_{-1}^0 dx_A \int_0^1 dx_B 
\langle \mathcal{O}(x_A) \otimes \mathcal{O}(x_B) \rangle , \\
N_{--} &=
N(x_A < 0, x_B < 0)
= \int_{-1}^0 dx_A \int_{-1}^0 dx_B 
\langle \mathcal{O}(x_A) \otimes \mathcal{O}(x_B) \rangle .
\end{align}
\end{subequations}
The notation $N(x_A > 0, x_B > 0)$ is a shorthand indicating that in practice one counts the events for which $x_A > 0$ and $x_B > 0$, and likewise for the other cases.

In terms of the Fano coefficients Eq.~\eqref{eq:encoding-simple-gen} becomes
\begin{subequations}
\label{eq:encoding-simple-fano}
\begin{align}
N_{++} &= \frac{1}{16}\Bigl( 4 + 2 B^+_x 
 + 2 B^-_x +  C_{xx} \Bigr), \\
N_{+-} &= \frac{1}{16}\Bigl( 4 + 2 B^+_x 
 - 2 B^-_x -  C_{xx} \Bigr), \\
N_{-+} &= \frac{1}{16}\Bigl( 4 - 2 B^+_x 
 + 2 B^-_x -  C_{xx} \Bigr), \\
N_{--}  &= \frac{1}{16}\Bigl( 4 - 2 B^+_x 
 - 2 B^-_x + C_{xx} \Bigr).
\end{align}
\end{subequations}
This leads to an expectation value, as in Eq.~\eqref{eq:exp-value-exp} of the outcome of the experiment of
\begin{equation}
\label{eq:exp-simple}
E = \frac{1}{4}C_{xx}.
\end{equation}
The prefactor of $1/4$ is related to the probability of a measured outcome yielding the same reported outcome which is
\begin{equation}
\label{eq:prob-simple}
P_+ = \int_0^1 dx \langle \mathcal{O}(x) \rangle_{+ x}
= \frac{3}{4}.
\end{equation}
The connection between the prefactor, the value of $P_+$, and the CHSH inequality will be derived in the following section.

\subsubsection{The General Case}
\label{sec:imperfect-general}

For the general case, we consider the measurement operator
\begin{equation}
\label{eq:opM-general}
\mathcal{O}(\hat{x})_{\hat{n}} = 
\frac{1}{2} \Bigl(
\mathbb{I}_2 + f(\hat{x} \cdot \hat{n}) 
\: \hat{n} \cdot \vec{\sigma} \Bigr),
\end{equation}
which is a function of $\hat{x}$ measured in the direction $\hat{n}$.  Here, $f$ is an odd function of the scalar quantity $\hat{x} \cdot \hat{n}$.

\begin{figure} 
  \centering
  \includegraphics[width=0.4\textwidth]{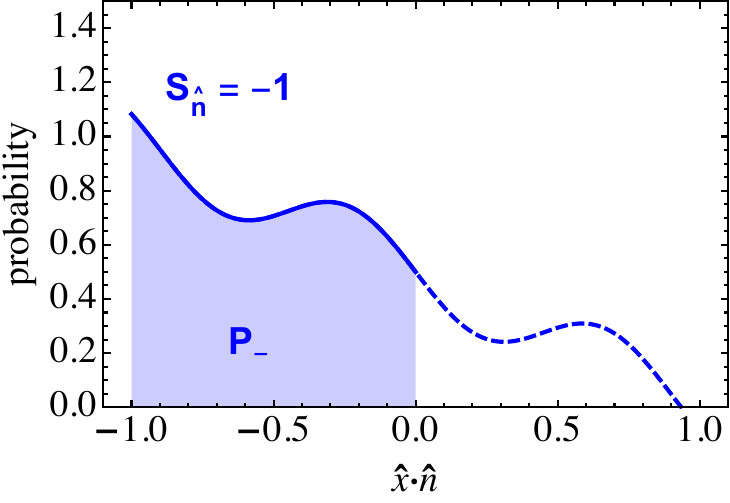}
  \qquad\qquad
    \includegraphics[width=0.4\textwidth]{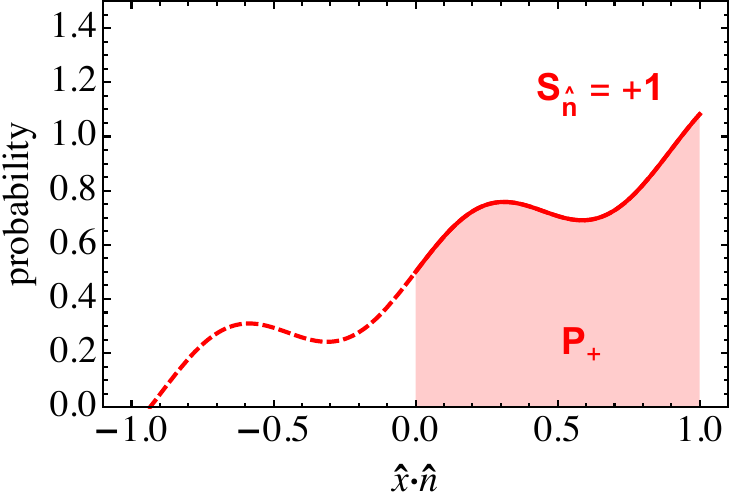}
  \caption{Distribution of reported outcomes from an imperfect detector for spin.  For underlying outcomes of $-$ the distribution is shown (left) and for underlying outcomes of $+$ the distribution is shown (right).}
  \label{fig:imperfect}
\end{figure}

The result of this measurement on an eigenstate $\rho_{\pm \hat{n}}$ is
\begin{equation}
\label{eq:exp-value-eigen-gen}
\langle \mathcal{O}(\hat{x})_{\hat{n}} \rangle_{\pm \hat{n}}
= \frac{1}{2}
\Bigl(
1 \pm f(\hat{x} \cdot \hat{n}) \Bigr).
\end{equation}
This mapping from underlying outcomes to reported outcomes is schematically shown in Fig.~\ref{fig:imperfect}.

The expectation value for one qubit is
\begin{equation}
\label{eq:exp-value-gen}
\langle \mathcal{O}(\hat{x})_{\hat{n}} \rangle
= \frac{1}{2}\Bigl( 1 + f(\hat{x} \cdot \hat{n}) \: \hat{n} \cdot \hat{B} \Bigr),
\end{equation}
and the expectation value for two qubits is
\begin{equation}
\label{eq:exp-value-gen-twoqubits}
\begin{aligned}
\langle \mathcal{O}(\hat{x}_A)_{\hat{\alpha}} \otimes \mathcal{O}(\hat{x}_B)_{\hat{\beta}} \rangle
=
\frac{1}{4}\Bigl(
1 + f(\hat{x}_A \cdot \hat{\alpha}) \: \hat{\alpha} \cdot \hat{B}^+
&+ f(\hat{x}_B \cdot \hat{\beta}) \: \hat{B}^- \cdot \hat{\beta} \\
&+ f(\hat{x}_A \cdot \hat{\alpha}) f(\hat{x}_B \cdot \hat{\beta}) 
\: \hat{\alpha} \cdot C \cdot \hat{\beta} \Bigr).
\end{aligned}
\end{equation}
The same encoding as in Eq.~\eqref{eq:encoding-simple-fano} can be used
\begin{subequations}
\label{eq:encoding-gen-gen}
\begin{align}
N_{++}(\hat{\alpha},\hat{\beta}) &= 
N(\hat{x}_A \cdot \hat{\alpha} > 0, \hat{x}_B \cdot \hat{\beta} > 0), \\
N_{+-}(\hat{\alpha},\hat{\beta}) &= 
N(\hat{x}_A \cdot \hat{\alpha} > 0, \hat{x}_B \cdot \hat{\beta} < 0), \\
N_{-+}(\hat{\alpha},\hat{\beta}) &= 
N(\hat{x}_A \cdot \hat{\alpha} < 0, \hat{x}_B \cdot \hat{\beta} > 0), \\
N_{--}(\hat{\alpha},\hat{\beta}) &=
N(\hat{x}_A \cdot \hat{\alpha} < 0, \hat{x}_B \cdot \hat{\beta} < 0),
\end{align}
\end{subequations}
and terms of the Fano coefficients
\begin{subequations}
\label{eq:encoding-gen-fano}
\begin{align}
N_{++}(\hat{\alpha},\hat{\beta}) &= \frac{1}{4}\Bigl( 1 + (\Delta F_A)(\hat{\alpha} \cdot \hat{B}^+) + (\Delta F_B)(\hat{\beta} \cdot \hat{B}^-) +   (\Delta F_A) (\Delta F_B) (\hat{\alpha} \cdot C \cdot \hat{\beta} ) \Bigr), \\
N_{+-}(\hat{\alpha},\hat{\beta}) &= \frac{1}{4}\Bigl( 1 + (\Delta F_A)(\hat{\alpha} \cdot \hat{B}^+) - (\Delta F_B)(\hat{\beta} \cdot \hat{B}^-) -   (\Delta F_A) (\Delta F_B) (\hat{\alpha} \cdot C \cdot \hat{\beta} ) \Bigr), \\
N_{-+}(\hat{\alpha},\hat{\beta}) &= \frac{1}{4}\Bigl( 1 - (\Delta F_A)(\hat{\alpha} \cdot \hat{B}^+) + (\Delta F_B)(\hat{\beta} \cdot \hat{B}^-) -   (\Delta F_A) (\Delta F_B) (\hat{\alpha} \cdot C \cdot \hat{\beta} ) \Bigr), \\
N_{--}(\hat{\alpha},\hat{\beta})  &= \frac{1}{4}\Bigl( 1 - (\Delta F_A)(\hat{\alpha} \cdot \hat{B}^+) - (\Delta F_B)(\hat{\beta} \cdot \hat{B}^-) +   (\Delta F_A) (\Delta F_B) (\hat{\alpha} \cdot C \cdot \hat{\beta} ) \Bigr).
\end{align}
\end{subequations}
The quantity $\Delta F_A$ is the integral of $f(\hat{x}_A \cdot \hat{\alpha})$ from $0$ to $1$ and $-\Delta F_A$ is the integral of $f(\hat{x}_A \cdot \hat{\alpha})$ from $-1$ to $0$.  The quantity $\Delta F_B$ is defined similarly.

The expectation value of the outcome of the measurement is
\begin{equation}
\label{eq:exp-gen}
E(\hat{\alpha},\hat{\beta})
=
(\Delta F_A \Delta F_B) \:
(\hat{\alpha} \cdot C \cdot \hat{\beta}).
\end{equation}
Meanwhile, the probability $P_+$ can be calculated given the encoding in Eq.~\eqref{eq:encoding-gen-gen}
\begin{equation}
\label{eq:prob-gen}
P_+ = \int_0^1 d(\hat{x} \cdot \hat{n}) \langle \mathcal{O}(\hat{x} \cdot \hat{n})_{\hat{n}} \rangle_{+ \hat{n}}
= \frac{1}{2}(1 + \Delta F),
\end{equation}
Finally, using Eq.~\eqref{eq:exp-gen} in the CHSH inequality leads to
\begin{equation}
\label{eq:chsh-general}
\epsilon_{\rm CHSH} \; \bigg|
\hat{\alpha}_1 \cdot C \cdot \hat{\beta}_1 
- \hat{\alpha}_1 \cdot C \cdot \hat{\beta}_2
+ \hat{\alpha}_2 \cdot C \cdot \hat{\beta}_1 +
\hat{\alpha}_2 \cdot C \cdot \hat{\beta}_2
\bigg|
\leq 2.
\end{equation}
Unlike the case of an ideal detector, here there is a theoretical efficiency factor $\epsilon_{\rm CHSH} = (\Delta F_A)(\Delta F_B)$ that dilutes the potential of a state to violate the inequality.\footnote{Ref.~\cite{Clauser:1978ng} called the theoretical efficiency factor $C$ rather than $\epsilon_{\rm CHSH}$.  }  This is not an experimental efficiency that can be corrected, but rather a theoretical efficiency that results from the measurement operator.  From Eq.~\eqref{eq:prob-gen} we find
\begin{equation}
\label{eq:eff-factor}
\epsilon_{\rm CHSH} = 
(2P_+^A - 1)(2P_+^B - 1).
\end{equation}
As the maximum value of the left-hand side for a Bell state is $2\sqrt{2}$ this means that Bell inequality violation is only observable in this system if $\epsilon_{\rm CHSH} > 1/\sqrt{2}$.  For $P_+ = P_+^A = P_+^B$ this results in the requirement $P_+ > 2^{-1}+2^{-5/4} = 0.92$.

From the previous simple example, we see that the value of $P_+^A = 3/4$ in Eq.~\eqref{eq:prob-simple} leads to the prefactor of $1/4$ in Eq.~\eqref{eq:exp-simple}.  These yield a theoretical efficiency factor of $\epsilon_{\rm CHSH} = 1/4$ which means that the CHSH inequality cannot be violated in such a system and consequently that local realism cannot be addressed.

\section{The CHSH Inequality at Colliders}
\label{sec:colliders}

The scenarios presented in Sec.~\ref{sec:detection} correspond to the detection scenarios at high-energy colliders under different assumptions.

\subsection{Assuming Spins}
\label{sec:collider-spin}

The first variation of a collider Bell test is to assume that the observed distributions are generated by underlying spins.  It is assumed, therefore, that the system is described by Eq.~\eqref{eq:fano-bloch} and that observed events are reconstructing the density matrix $\rho$ by means of measuring the Fano-Bloch coefficients.  This corresponds to the ideal detector scenario of Sec.~\ref{sec:ideal}. 

With this assumption, an observed spin can have the outcomes $+1$ or $-1$, statistically reconstructed from data, leading to the direct application of Eq.~\eqref{eq:CHSH} via Eq.~\eqref{eq:chsh-spins}.\footnote{Generally, the states reconstructed at colliders are fictitious states rather than quantum states~\cite{Afik:2022kwm,Cheng:2023qmz,Cheng:2024btk}.  The distinction here is not critical because when a fictitious state is Bell nonlocal its corresponding quantum state is also Bell nonlocal~\cite{Afik:2022kwm,Cheng:2023qmz,Cheng:2024btk}.}

The expectation value $E(\vec{\alpha},\vec{\beta})$ is constructed from the appropriate reconstructed set of Fano-coefficients
\begin{equation}
\label{eq:exp-collider-spins}
E(\vec{\alpha},\vec{\beta}) = \vec{\alpha} \cdot C \cdot \vec{\beta}.
\end{equation}
As discussed in Sec.~\ref{sec:Bell}, measuring Bell's inequality is fundamentally a test of local realism. However, by assuming an underlying spin structure, one effectively builds the violation of local realism into the model itself. As a result, this variation cannot serve as a meaningful test of local realism.

However, assuming spin does not invalidate the presence of quantum correlations. This still constitutes a meaningful measurement of Bell nonlocality.  This approach to measuring the CHSH inequality is the one currently used in all recent studies, implying that while current detectors at high-energy colliders can probe the correlations implied by Bell nonlocality, they cannot test local realism.

\subsection{Without Assuming Spins}
\label{sec:collider-angle}

The second variation of a collider Bell test does not make any assumptions about the underlying description.  This variation corresponds to the scenario of an imperfect detector described in Sec.~\ref{sec:imperfect}.\footnote{This situation in collider physics of whether or not spins are assumed has an analogy in quantum cryptography~\cite{Scarani:2009zz}.  In quantum cryptography when security requires trusting Alice and Bob's devices the protocol is called device-dependent quantum key distribution.  When the system used is entangled but cannot be used to disprove local realism, trusting the devices is necessary.  When a quantum system used for quantum cryptography can demonstrate the exclusion of local realism, then it is no longer necessary for Alice and Bob to trust their devices.  This protocol is called device-independent quantum key distribution~\cite{Ekert:1991zz,Mayers:2003xif,Ac_n_2007}.}

The observed outcomes are rest-frame decay-product angles.  Consider the production of two particles $A$ and $B$.  We are interested in the correlation between the spin of $A$ and the spin of $B$.  If the particle $A$ has an $n$-body decay, according to $A \to a, a_2, \ldots, a_n$, we consider the particle $a$, without loss of generality, to be the spin analyzer.  The spin analyzing power is the coefficient $\kappa_a$ in the measurement operator
\begin{equation}
\label{eq:opM-spin-power}
\mathcal{O}(\hat{p}_a)
= \frac{1}{2}\left( 1 + \kappa_a \hat{p}_a \cdot \vec{\sigma} \right).
\end{equation}
Along the direction $\hat{n}$ this operator is
\begin{equation}
\label{eq:opM-spin-power-gen}
\mathcal{O}(\hat{p}_a)_{\hat{n}}
= \frac{1}{2}\left( 1 + \kappa_a (\hat{p}_a \cdot \hat{n}) \hat{n} \cdot \vec{\sigma} \right),
\end{equation}
Taking the expectation value in the direction $\hat{n}$ leads to
\begin{equation}
\label{eq:exp-val-eigen-spin-power}
\langle \mathcal{O}(\hat{p}_a) \rangle_{\pm \hat{n}}
= \frac{1}{2}\left( 1 \pm \kappa_a \cos\theta_{a,\hat{n}}\right),
\end{equation}
where $\theta_a$ is the angle between $\hat{n}$ and the normalized three-momentum of $a$, $\hat{p}_a$, in the rest frame of $A$.  Similarly, the spin analyzing power $\kappa_b$ is defined for the decay $B \to b, b_2, \ldots, b_n$, with $\hat{p}_b$ denoting the normalized three-momentum of $b$ in the rest frame of $B$.  Eq.~\eqref{eq:exp-val-eigen-spin-power} is also the normalized differential decay width of $A$.

\begin{figure} 
  \centering
    \includegraphics[width=0.4\textwidth]{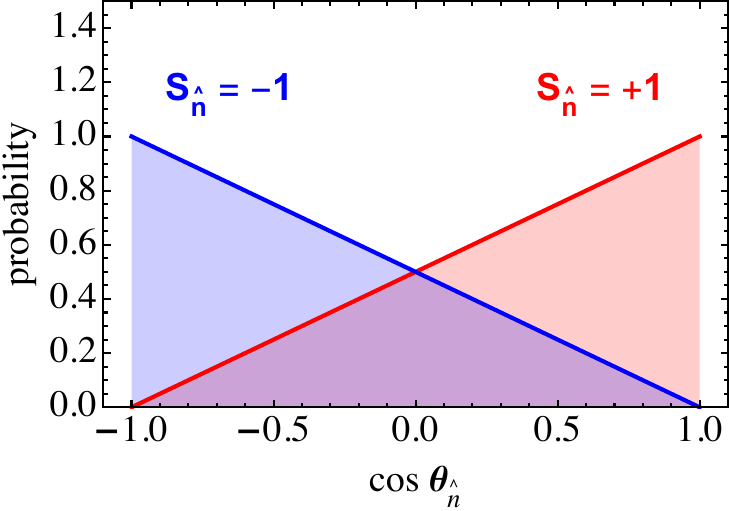}
  \caption{Distribution of reported outcomes of spin along the $i$ axis for spin $=-1$ (blue) and spin $=+1$ (red) with a direct spin measurement (left) and using the rest-frame decay-product angle (right), assuming $\kappa_a=\kappa_b=1$.}
  \label{fig:dist-spin}
\end{figure}

This corresponds to the detection function in Eq.~\eqref{eq:opM-general} with $f(\hat{x}_a \cdot \hat{n}) = \kappa_a \cos\theta_{a,\hat{n}}$ and is shown in Fig.~\ref{fig:dist-spin}.  For simplicity, in this work we pretend that all $\kappa$ are positive.  To accommodate negative values, one needs to swap the integration regions between $[0,1]$ and $[-1,0]$ in the appropriate places.

From Eq.~\eqref{eq:opM-spin-power} we find that $P_+ = (1/2)(1 + \kappa/2)$.  We see intuitively that when $\kappa = 0$, corresponding to no spin information transfer between the particle spin and the rest-frame decay-product angle, resulting in $\epsilon_{\rm CHSH} \to 0$ meaning that there is no possibility to violate Bell's inequality.  Generally, we find
\begin{equation}
\label{eq:eff-factor-collider}
\epsilon_{\rm CHSH} = \frac{\kappa_a \kappa_b}{4},
\end{equation}
which means that even for $\kappa_a = \kappa_b = 1$ the level of correlation between the underlying spin and the observed rest-frame decay-product angle is insufficient to allow a violation of Bell's inequality.

\subsection{Without External Detector Settings}
\label{sec:collider-kasday}

In fact, the result that a test of local realism cannot be performed at colliders was already foreseeable from the fact that the rates in Eq.~\eqref{eq:encoding-gen-gen} can all be expressed as cross sections.  Results expressed as cross sections can be described by the Kasday's hidden variable model~\cite{Abel:1992kz,Dreiner:1992gt,Bechtle:2025ugc,Abel:2025skj}.

Another way to view the issue is the following.  Refs.~\cite{Abel:1992kz,Dreiner:1992gt,Bechtle:2025ugc,Abel:2025skj} argued that when spins are assumed, one has the three-vector $\vec{S}_A$, $\vec{S}_B$, $\vec{\alpha}$, and $\vec{\beta}$, which are the spin of $A$, the spin of $B$, the measurement axis for $A$, and the measurement axis for $B$, respectively.  This allows for invariants like $\vec{S}_A \cdot \vec{\alpha}$ which do not allow the expectation values to be expressed as a differential cross section.  When spins are not assumed, the only three-vectors available are $\vec{p}_a$ and $\vec{p}_b$ which only allows the invariant $\vec{p}_a \cdot \vec{p}_b$.  

Our work, in Sec.~\ref{sec:imperfect}, effectively uses the three-vectors $\vec{p}_a$, $\vec{p}_b$, $\vec{\alpha}$, and $\vec{\beta}$.  While this allows invariants such as $\vec{p}_a \cdot \vec{\alpha}$, unlike when spins are assumed, these are still expressible as a differential cross section.  The reason is that $\vec{\alpha}$ and $\vec{\beta}$ are not independently chosen when applied to rest-frame decay-product angles.  Due to Lorentz invariance, these reference directions, are part of the system itself and subsequently can still be expressed as a differential cross section.

\subsection{At Future Detectors}
\label{sec:collider-future}

It is possible that at future collider experiments the associated detectors will be able to measure the spins of particles~\cite{Doser:2022knm,Afik:2025ejh}.  If the spin measurements are direct, as in Sec.~\ref{sec:ideal}, then the experimenter would choose a measurement axis.  In this case, Kasday's local hidden variable model does not apply and local realism can be tested, when the measurement axes can be changed as needed.

It is not necessary to measure spins on an event-by-event basis.  Since quantum mechanics is fundamentally probabilistic, spins can be inferred statistically.  In this case, testing local realism is still possible provided that $\epsilon_{\rm CHSH} > 1/\sqrt{2}$ or equivalently $P_+ > 0.92$, when $P_+ = P_+^A = P_+^B$.

\subsection{Beyond Quantum Mechanics}
\label{sec:beyondQM}

The CHSH inequality in its dual-channel form, as given in Eq.~\eqref{eq:CHSH}, can be expressed as
\begin{equation} 
\label{eq:s-chsh}
S_{\rm CHSH} \leq 2,
\end{equation}
using the quantity $S_{\rm CHSH}$
\begin{equation}
\label{eq:chsh-with-s}
S_{\rm CHSH} = |E(\alpha_1,\beta_1) 
- E(\alpha_1,\beta_2) 
+ E(\alpha_2,\beta_1) 
+ E(\alpha_2,\beta_2)|.
\end{equation}
Given the range spanned by the expectation value $E(\alpha,\beta)$, the algebraic limit of $S_{\rm CHSH}$ is 4.  However, in theories that respect local realism,  the tighter bound of Eq.~\eqref{eq:s-chsh} applies.

Quantum mechanical theories allow for a violation of the CHSH inequality but themselves are bounded by
\begin{equation}
\label{eq:cirelson}
S_{\rm CHSH} \leq 2 \sqrt{2},
\qquad\qquad\qquad
\text{(Cirelson bound)},
\end{equation}
which is known as the Cirelson bound~\cite{Cirelson:1980ry}.  The class of theories that can exceed the Cirelson bound are called no-signaling theories and they, in turn, are bounded by the Popescu-Rohrlich (PR) box~\cite{Popescu:1994kjy,Popescu:2014wva}.  For $S_{\rm CHSH}$ the PR box corresponds to the algebraic limit
\begin{equation}
\label{eq:PR-box}
S_{\rm CHSH} \leq 4,
\qquad\qquad\qquad
\text{(PR box)}.
\end{equation}
For single-channel setups, as in Eq.~\eqref{eq:CH1974}, we have
\begin{equation}
\label{eq:sch}
-1 \leq S_{\rm CH} \leq 0,
\end{equation}
using the quantity $S_{\rm CH}$
\begin{equation}
\label{eq:ch-with-s}
S_{\rm CH} = R(a_1, b_1) - R(a_1,b_2) + R(a_2,b_1) + R(a_2,b_2) - R(a_2,\infty) - R(\infty,b_1).
\end{equation}
The corresponding Cirelson bound is
\begin{equation}
\label{eq:cirelson-ch}
-\frac{1}{2} - \frac{1}{\sqrt{2}} \leq S_{\rm CH} \leq - \frac{1}{2} + \frac{1}{\sqrt{2}},
\qquad\qquad\qquad
\text{(Cirelson bound)},
\end{equation}
and the corresponding PR box is
\begin{equation}
\label{eq:PR-box-ch}
-\frac{3}{2} \leq S_{\rm CH} \leq \frac{1}{2},
\qquad\qquad\qquad
\text{(PR box)}.
\end{equation}
In addition to testing local realism one could attempt to test for the presence of superquantum correlations that violate the Cirelson bound but lie within the PR box.  We show that with the current analysis techniques, this is not possible at colliders.

The ideal test would not assume spins and utilize rest-frame decay-product angles directly, as presented in Sec.~\ref{sec:collider-angle}.  The theoretical efficiency factor $\epsilon_{\rm CHSH}$ enters Eq.~\eqref{eq:cirelson} via $S_{\rm CHSH} \to \epsilon_{\rm CHSH} S_{\rm CHSH}$ which renders the Cirelson bound untestable.  Unlike with local realism, while testing the Cirelson bound it is not invalidating to assume the underlying description is built on spins.  This case, however, is also untestable because of the measurement method applied at colliders.

To perform quantum tomography at colliders one extracts the Fano coefficients via fits to rest-frame angular decay distributions.  As each Fano coefficient is related to a differential cross section they are each bounded to have an absolute value less than or equal to 1.  A properly reconstructed quantum density matrix will have all non-negative eigenvalues.  The measurement process at a collider allows the reconstruction of density matrices that have one or more negative eigenvalues which does not correspond to a valid quantum state.  Cirelson's proof requires that the operator norm corresponding to each Fano coefficient be bounded between $-1$ and $1$ but has no requirement on the validity of the quantum state~\cite{Cirelson:1980ry}.   These invalid quantum states, therefore, as measured at colliders, cannot violate the Cirelson bound.

One possibility to indicate superquantum behavior would be deviations in the distributions used to extract the Fano coefficients.  Distributions that do not match the quantum field theory predictions, if verified, could point to a superquantum theory (see, for example, Ref.~\cite{Altomonte:2024upf}).


\section{Entanglement and Bell Nonlocality}
\label{sec:entanglement}

As noted in Refs.~\cite{Abel:1992kz,Dreiner:1992gt,Bechtle:2025ugc,Abel:2025skj}, local realism cannot be tested in collider experiments because only cross sections are measured, allowing a local hidden variable model to account for all observed data.\footnote{Refs.~\cite{Abel:1992kz,Dreiner:1992gt,Bechtle:2025ugc,Abel:2025skj} also state that the reason cross sections have a local hidden variable model is that the measured final state momenta commute.  It has been shown that commuting measurements can still lead to tests of local realism if the choice of detector settings was made at some point in the past and transmitted, along with the results, to the experimenters~\cite{Maldacena:2015bha}.}  In Sec.~\ref{sec:detection} and Sec.~\ref{sec:colliders} we have demonstrated this in an alternative way, namely that due to the distribution of decay products, the theoretical efficiency factor multiplying the CHSH inequality is too low to test local realism.

In Refs.~\cite{Abel:1992kz,Dreiner:1992gt,Abel:2025skj} the chosen expectation value, to be used in Bell's inequality, is
\begin{equation}
\label{eq:dreiner-exp-xsec}
E(A,B) = 
\frac{1}{\sigma(IJ \to AB)}
\frac{d\sigma(IJ \to AB)}{d\cos \theta_{ab}},
\end{equation}
where $I$ and $J$ are incoming particles and $A$ and $B$ are outgoing particles.  This is parametrized as
\begin{equation}
\label{eq:dreiner-exp-entanglement}
E(A,B) = \frac{1}{2}\left( 1 - D \cos\theta_{ab} \right),
\end{equation}
where $\theta_{ab} = \hat{p}_a \cdot \hat{p}_b$ where $\hat{p}_A$ and $\hat{p}_B$ are the normalized momenta, respectively, of particles $a$ and $b$.  Particle $a$ is one of the decay products of particle $A$ and particle $b$ is one of the decay products of particle $B$.  The parameter $D$ is related to the Fano coefficients from Eq.~\eqref{eq:fano-bloch} via the trace of the spin correlation matrix $D = {\rm tr}(C)/3$ and has a value that is determined by the process. 

Refs.~\cite{Abel:1992kz,Abel:2025skj} apply the expectation value to Bell's inequality through Bell's original inequality in Eq.~\eqref{eq:bell-original}.  This inequality, however, only applies to spins in a singlet-configuration, meaning that their expectation value is fully anti-correlated, which is not the spin state of the $e^+ e^- \to Z \to \tau^+ \tau^-$ process~\cite{Tsai:1971vv}.  The CHSH inequality, in Eq.~\eqref{eq:CHSH}, applies to any spin configuration.  Ref.~\cite{Dreiner:1992gt} does use the CHSH inequality and claims that the inequality is never violated for $|D| \leq 1$.  In contrast, we find using the methods of Ref.~\cite{Dreiner:1992gt} that the inequality is never violated for $|D| \leq 1/2$.  Since there are known Bell nonlocal states with $|D| > 1/2$~\cite{Horodecki:1996qk}, the example from Ref.~\cite{Dreiner:1992gt} is sufficient for $e^+ e^- \to Z \to \tau^+ \tau^-$, but insufficient for other processes.

Using the CHSH inequality, however, as discussed in Sec.~\ref{sec:Bell}, requires an appropriate choice of expectation value which should span positive and negative values.  Eq.~\eqref{eq:dreiner-exp-entanglement} spans the values from $(1-|D|)/2$ to $(1+|D|)/2$.  Given the inefficiency of this expectation value, Refs.~\cite{Abel:1992kz,Dreiner:1992gt,Abel:2025skj} do not leverage the correlations between terms as is typical for a CHSH bound but rather take an algebraic limit.
In order to use Eq.~\eqref{eq:dreiner-exp-entanglement} either one should use the CH inequality from Eq.~\eqref{eq:CH1974} (as we show in Appendix~\ref{app:encoding}) or select an encoding for $+1$ and $-1$ (as we do in Sec.~\ref{sec:imperfect-general}). Ref.~\cite{Bechtle:2025ugc} does utilize the correlations, but finds a less efficient bound than we find.

The parameter $D$ is often useful for studying entanglement.  The presence of entanglement can be shown by measuring the concurrence~\cite{Wootters:1997id}, which is an entanglement monotone~\cite{Horodecki:2009zz}, of a quantum state.  In some phase space regions of some final states, like $t\bar{t}$ near threshold, the parameter $D$ is equal to the concurrence~\cite{Afik:2020onf}.  Ref.~\cite{Bechtle:2025ugc} points out that the concurrence measured by ATLAS and CMS cannot be model-independently claimed to be non-zero.  This is because extracting the spin information from the rest-frame decay-product angular distributions assumes the Standard Model including the values of the spin analyzing powers.

Ref.~\cite{Abel:2025skj} incorrectly links the existence of a local hidden variable model that describes the quantum states at colliders with the possibility of testing entanglement vs. non-entanglement.  It is known that entangled states that are Bell local have a description via a local hidden variable model~\cite{Werner:1989zz} (see Ref.~\cite{Augusiak_2014} for a review).  Entanglement means that a state cannot be fully described by its individual components, meaning that it is not separable, and by itself makes no statement about local hidden variable models.

The ATLAS~\cite{ATLAS:2023fsd} and CMS~\cite{CMS:2024pts,CMS:2024zkc} experiments have validly measured the entanglement, via the concurrence, of the $t\bar{t}$ quantum state, assuming the Standard Model.  CMS~\cite{CMS:2024zkc} has tested non-zero entanglement versus zero entanglement, assuming the Standard Model values of the spin analyzing powers.

\section{Conclusions}
\label{sec:conclusions}

Quantum information theory applied to high-energy collider physics has led to the study of a variety of new quantum systems.  Using the spins of outgoing particles as the fundamental quantum unit, the qubits, has been very fruitful because these spins are often produced with non-classical correlations.

These quantum correlations between particle spins allow for the classification into classes of quantum states such as entangled states, Bell nonlocal states, and many others.  These classifications indicate the amount of quantum resources present in such states.  Bell nonlocality is particularly significant among quantum correlations because, under suitable assumptions, it marks the boundary beyond which local realism can no longer describe nature.

Whether high-energy colliders are valid tests of local realism has been a recent topic of interest.  In this work, we addressed this question directly by calculating how a Bell test would be performed at a high-energy collider.  We showed that it is not possible to exclude local realism at a collider because the observables used, rest-frame decay-product angles, are not sufficiently correlated with the underlying particle spins.

This was captured by a theoretical efficiency factor that dilutes the observation of the terms in the CHSH inequality relative to the threshold required for violation.  We showed that for a Bell test to be possible the theoretical efficiency factor must be greater than $1/\sqrt{2}$.  At high-energy colliders, the largest possible factor is $1/4$, rendering this an ineffective test.  Future detectors that could measure particle spins directly would be able to test local realism because experiments in which spin is directly measured have a theoretical efficiency factor of $1$.

Despite this conclusion, Bell nonlocality, as a quantum correlation, remains an interesting and informative threshold for correlations at colliders.

\begin{acknowledgments}
ML thanks Kun Cheng and Tao Han for detailed feedback and thanks Yoav Afik, Marco Fabbrichesi, Ian Low, Juan Maldacena, Luca Marzola, Juan Ram\'on Mu\~noz de Nova, Kazuki Sakurai, and Xerxes Tata for helpful discussions.  ML is supported by the US Department of Energy under grant No. DE-SC0007914, by the National Science Foundation under grant No. PHY-2412696, and by Pitt PACC.
\end{acknowledgments}

\appendix
\section{Alternative Encoding}
\label{app:encoding}

In the main text, we presented encodings that describe how to map measured continuous values to the reported outcomes of $+1$ and $-1$.  Given these encodings the CHSH inequality from Eq.~\eqref{eq:CHSH} can be used.  An intuitive alternative to the encodings is to use the event rates directly in the CH inequality from Eq.~\eqref{eq:CH1974}.  In this appendix, we show that this approach gives identical results.

Consider the general case shown in Sec.~\ref{sec:imperfect-general}, with the measurement operator from Eq.~\eqref{eq:opM-general}
\begin{equation*}
\mathcal{O}(\hat{x})_{\hat{n}} = 
\frac{1}{2} \bigl(
\mathbb{I}_2 + f(\hat{x} \cdot \hat{n}) \hat{n} \cdot \vec{\sigma} \bigr).
\end{equation*}
After choosing two directions $\hat{\alpha}$ and $\hat{\beta}$ the rates are given by
\begin{subequations}
\label{eq:alt-encoding-gen}
\begin{align}
R(\hat{\alpha}, \hat{\beta}) &= 
\frac{N(\hat{x}_A \cdot \hat{\alpha} > 0, \hat{x}_B \cdot \hat{\beta} > 0)}{N_{\rm tot}}, \\
R(\hat{\alpha},\infty) &=
\frac{N(\hat{x}_A \cdot \hat{\alpha} > 0)}{N_{\rm tot}}, \\
R(\infty, \hat{\beta}) &=
\frac{N(\hat{x}_B \cdot \hat{\beta} > 0)}{N_{\rm tot}},
\end{align}
\end{subequations}
where $N_{\rm tot}$ is the number of events with no restriction on either qubit $A$ or qubit $B$. 

In terms of the Fano coefficients, the rates are
\begin{subequations}
\label{eq:alt-encoding-fano}
\begin{align}
R(\hat{\alpha},\hat{\beta}) &=
\frac{1}{4}\left(1 + (\hat{\alpha} \cdot \hat{B}^+) (\Delta F_A) + (\hat{B}^- \cdot \hat{\beta}) (\Delta F_B) + (\hat{\alpha} \cdot C  \cdot \hat{\beta}) (\Delta F_A) (\Delta F_B) \right), \\
R(\hat{\alpha},\infty) &=
\frac{1}{2}\left(1 + (\hat{\alpha} \cdot \hat{B}^+) (\Delta F_A) \right), \\
R(\infty,\hat{\beta}) &=
\frac{1}{2}\left(1 + (\hat{B}^- \cdot \hat{\beta}) (\Delta F_B) \right).
\end{align}
\end{subequations}
The quantity $\Delta F_A$ is the integral of $f(\hat{x}_A \cdot \hat{\alpha})$ from $0$ to $1$ and $-\Delta F_A$ is the integral of $f(\hat{x}_A \cdot \hat{\alpha})$ from $-1$ to $0$.  The quantity $\Delta F_B$ is defined similarly.

The resulting CH equation is
\begin{equation}
\label{eq:ch-fano}
\begin{aligned}
-1 \leq \frac{1}{4}
\bigg( -2 + (\Delta F_A)(\Delta F_B) (
\hat{\alpha}_1 \cdot C \cdot \hat{\beta}_1 & - \hat{\alpha}_1 \cdot C \cdot \hat{\beta}_2  \\
& + \hat{\alpha}_2 \cdot C \cdot \hat{\beta}_1 + \hat{\alpha}_2 \cdot C \cdot \hat{\beta}_2
)
\bigg) \leq 0,
\end{aligned}
\end{equation}
which is equivalent to Eq.~\eqref{eq:chsh-general}
\begin{equation*}
\epsilon_{\rm CHSH} \; \bigg|
\hat{\alpha}_1 \cdot C \cdot \hat{\beta}_1 
- \hat{\alpha}_1 \cdot C \cdot \hat{\beta}_2
+ \hat{\alpha}_2 \cdot C \cdot \hat{\beta}_1 +
\hat{\alpha}_2 \cdot C \cdot \hat{\beta}_2
\bigg|
\leq 2,
\end{equation*}
for $\epsilon_{\rm CHSH} = \Delta F_A \Delta F_B$.

For colliders, the measurement operator is the decay operator which leads to the form
\begin{equation*}
\mathcal{O}(\hat{p})_{\hat{n}} = 
\frac{1}{2} \bigl(
\mathbb{I}_2 + \kappa (\hat{p} \cdot \hat{n}) \hat{n} \cdot \vec{\sigma} \bigr)
=
\frac{1}{2} \bigl(
\mathbb{I}_2 + \kappa \cos\theta_{\hat{n}} \hat{n} \cdot \vec{\sigma} \bigr).
\end{equation*}
The rates can be directly calculated from the doubly-differential cross section
\begin{subequations}
\label{eq:alt-encoding-xsec}
\begin{align}
R(\hat{\alpha},\hat{\beta}) 
&= \frac{1}{\sigma} 
\int_{0}^{1} d\cos\theta_{a,\hat{\alpha}} 
\int_{0}^{1} d\cos\theta_{b,\hat{\beta}}
\left(\frac{d^2\sigma}{d\cos\theta_{a,\hat{\alpha}} d\cos\theta_{b,\hat{\beta}}}\right), \\
R(\hat{\alpha},\infty) 
&= \frac{1}{\sigma} 
\int_{0}^{1} d\cos\theta_{a,\hat{\alpha}}
\int_{-1}^{1} d\cos\theta_{b,\hat{\beta}}
\left(\frac{d^2\sigma}{d\cos\theta_{a,\hat{\alpha}} d\cos\theta_{b,\hat{\beta}}}\right), \\
R(\infty,\hat{\beta}) 
&= \frac{1}{\sigma} 
\int_{-1}^{1} d\cos\theta_{a,\hat{\alpha}} 
\int_{0}^{1} d\cos\theta_{b,\hat{\beta}}
\left(\frac{d^2\sigma}{d\cos\theta_{a,\hat{\alpha}} d\cos\theta_{b,\hat{\beta}}}\right),
\end{align}
\end{subequations}
and in terms of the Fano coefficients, they are
\begin{subequations}
\label{eq:alt-encoding-collider}
\begin{align}
R(\hat{\alpha},\hat{\beta}) 
&= 
\frac{1}{4} 
+ \frac{\kappa_a}{8} \hat{\alpha} \cdot \hat{B}^+ + \frac{\kappa_b}{8} \hat{B}^- \cdot \hat{\beta} 
+ \frac{\kappa_a \kappa_b}{16} \hat{\alpha} \cdot C \cdot \hat{\beta} , \\
R(\hat{\alpha},\infty) 
&= \frac{1}{2} 
+ \frac{\kappa_a}{4} \hat{\alpha} \cdot \hat{B}^+, \\
R(\infty,\hat{\beta}) 
&= \frac{1}{2} 
+ \frac{\kappa_b}{4} \hat{B}^- \cdot \hat{\beta}.
\end{align}
\end{subequations}

\bibliographystyle{utphys}
\bibliography{refs,refs_collider}
\end{document}